\documentclass[sigconf, nonacm]{acmart}
 
\newcommand\vldbdoi{XX.XX/XXX.XX}
\newcommand\vldbpages{XXX-XXX}
\newcommand\vldbvolume{14}
\newcommand\vldbissue{1}
\newcommand\vldbyear{2020}
\newcommand\vldbauthors{Manos Chatzakis, Rodrigo Flores-Allende, Mikael Freire, Yoann Cano, L\'eonard Seydoux, and Themis Palpanas}
\newcommand\vldbtitle{\shorttitle} 
\newcommand\vldbavailabilityurl{URL_TO_YOUR_ARTIFACTS}
\newcommand\vldbpagestyle{plain} 

\usepackage{enumitem}
\usepackage[export]{adjustbox}
\usepackage{xcolor}
\usepackage{booktabs}
\usepackage{subcaption}
\usepackage{hyperref}
\usepackage{amsmath}
\usepackage[vlined,ruled,linesnumbered,noend]{algorithm2e}
\usepackage{cleveref}
\usepackage[normalem]{ulem}
\usepackage{listings}
\usepackage{graphicx}
\usepackage{caption}
\usepackage{adjustbox}
\usepackage{amsmath}
\usepackage{tikz}
\usetikzlibrary{tikzmark}
\usepackage{mathtools}

\newcommand{\todo}[1]{\textcolor{red}{#1}}
\begin{document}

\title{TREMOR: Template Matching for Large Seismic Data Collections}

%\makeatletter\def\@authorfont{\large}\makeatother % smaller author font, so that all names fit in one line
\author{Manos Chatzakis$^{1}$, Rodrigo Flores-Allende$^{2}$, Mikael Freire$^{3}$, Yoann Cano$^{3}$, L\'eonard Seydoux$^{2}$, Themis Palpanas$^{1}$}
\affiliation[obeypunctuation=true]{%
  \institution{$^{1}$Universit\'e Paris Cit\'e, LIPADE, $^{2}$Universit\'e Paris Cit\'e, IPGP, $^{3}$CEA, DAM, DIF}\\
  {\normalfont manos.chatzaki@gmail.com, floresallende@ipgp.fr, mikael.freire@cea.fr, yoann.cano@cea.fr, seydoux@ipgp.fr, themis@mi.parisdescartes.fr}
}

% OLD author block (one block per author):
% \author{Manos Chatzakis}
% \email{manos.chatzaki@gmail.com}
% \affiliation{
%   \institution{Universit\'e Paris Cit\'e, LIPADE}
%   %\city{F-75006 Paris}
%   %\country{France}
% }

% \author{Rodrigo Flores-Allende}
% \email{floresallende@ipgp.fr}
% \affiliation{
%   \institution{Université Paris Cité, IPGP}
%   %\city{F-75005 Paris}
%   %\country{France}
% }

% \author{Mikael Freire}
% \email{mikael.freire@cea.fr}
% \affiliation{
%   \institution{CEA, DAM, DIF}
%  % \city{F-91297 Arpajon}
%   %\country{France}
% }

% \author{Yoann Cano}
% \email{yoann.cano@cea.fr}
% \affiliation{
%   \institution{CEA, DAM, DIF}
%  % \city{F-91297 Arpajon}
%  % \country{France}
% }

% \author{Léonard Seydoux}
% \email{seydoux@ipgp.fr}
% \affiliation{
%   \institution{Université Paris Cité, IPGP}
%  % \city{F-75005 Paris}
%  % \country{France}
% }

% \author{Themis Palpanas}
% \email{themis@mi.parisdescartes.fr}
% \affiliation{
%   \institution{Universit\'e Paris Cit\'e, LIPADE}
%  % \city{F-75006 Paris}
%  % \country{France}
% }

\renewcommand{\shortauthors}{Chatzakis et al.}

\settopmatter{printacmref=false}

\begin{abstract}
Seismic station networks continuously record the ground velocity at several locations on Earth in the form of waveform data series, 
which seismologists analyze to detect various kinds of geophysical events, including earthquakes. 
%Because earthquakes that occur in the same context may produce very similar waveforms, 
Then, {\em template} events are defined to search the seismic data collection for {\em matching} events.
This is %known as 
{\em template matching}, a fundamental task in seismology %, serving as the backbone 
for various seismic analyses.
However, template matching requires extensive processing times, especially for large seismic collections that exceed the memory capacity of a single machine. 
This poses a significant challenge in seismology, in particular with the increasing size of seismological datasets.
In this paper, we introduce TREMOR, a distributed data series processing framework for template matching, that efficiently handles large waveform collections.
We apply TREMOR to two representative real-world seismic use cases. % for template matching 
Through an extensive experimental evaluation, we demonstrate its efficiency, with TREMOR being up to 6.7$\times$ faster than the best competing method on each workload, and up to 18$\times$ faster than the fastest distributed version of MASS, an FFT-based method popular in seismology, while returning the same, exact results.
\end{abstract}  

%\keywords{Data Series, Template Matching, Subsequence Similarity Search, Distributed Systems}

\maketitle

%%% do not modify the following VLDB block %%
%%% VLDB block start %%%
\pagestyle{\vldbpagestyle}
\begingroup\small\noindent\raggedright\textbf{PVLDB Reference Format:}\\
\vldbauthors. \vldbtitle. PVLDB, \vldbvolume(\vldbissue): \vldbpages, \vldbyear.\\
\href{https://doi.org/\vldbdoi}{doi:\vldbdoi}
\endgroup
\begingroup
\renewcommand\thefootnote{}\footnote{\noindent
This work is licensed under the Creative Commons BY-NC-ND 4.0 International License. Visit \url{https://creativecommons.org/licenses/by-nc-nd/4.0/} to view a copy of this license. For any use beyond those covered by this license, obtain permission by emailing \href{mailto:info@vldb.org}{info@vldb.org}. Copyright is held by the owner/author(s). Publication rights licensed to the VLDB Endowment. \\
\raggedright Proceedings of the VLDB Endowment, Vol. \vldbvolume, No. \vldbissue\ %
ISSN 2150-8097. \\
\href{https://doi.org/\vldbdoi}{doi:\vldbdoi} \\
}\addtocounter{footnote}{-1}\endgroup
%%% VLDB block end %%%

%%% do not modify the following VLDB block %%
%%% VLDB block start %%%
\ifdefempty{\vldbavailabilityurl}{}{
\vspace{.3cm}
\begingroup\small\noindent\raggedright\textbf{PVLDB Artifact Availability:}\\
The source code, data, and/or other artifacts have been made available at \url{https://github.com/MChatzakis/TREMOR}.
\endgroup
}
%%% VLDB block end %%%

\section{Introduction}
\label{sec:introduction}

Seismologists rely on {\em seismic stations}, which continuously monitor seismic activity worldwide~\cite{rost2002array,romanowicz2010seismology,havskov2016seismic}.
Seismic stations record ground velocity over time, producing {\em waveforms} represented as {\em data series}, i.e., long sequences of real-valued, time-ordered points~\cite{romanowicz2010seismology}.
Each station records ground motion along three {\em channels}, i.e., northward, eastward, and upward, and collections of stations form a {\em seismic network}~\cite{rost2002array,havskov2016seismic}.
Seismologists analyze these waveforms to detect seismic {\em events}, such as earthquakes, e.g.,~\cite{shelly2020high}, seismo-volcanic activity, e.g.,~\cite{journeau2020detection}, and other geophysical phenomena. % of interest.

Seismologists usually perform first-order detection using either statistical waveform features~\cite{baillard2014automatic} or deep learning tools~\cite{zhu2019phasenet}.
However, such approaches may miss small earthquakes when their signals are masked by background noise or overlapping events. 
Some can be recovered using waveforms from detected events as {\em templates}, identifying similar earthquakes without independently picking seismic phases~\cite{gibbons2006detection,shelly2007non,peng2009migration}. 
Detection can use individual stations or multiple stations jointly to improve accuracy, known as seismic {\em array processing}~\cite{rost2002arrayprocessing}. Templates are time series orders of magnitude shorter than continuous recordings. 
Seismic {\em template matching}, a fundamental process in seismology, searches these recordings for similar {\em subsequences} using a {\em sliding window}, identifying either the {\em $k$-nearest} matches or those exceeding a {\em similarity threshold}~\cite{frank2012slidingwindowtemplatematching}. 
Similarity is quantified using measures such as {\em Euclidean distance} or {\em cross correlation coefficient}~\cite{shelly2007non}.

\noindent \textbf{Motivating Applications and Challenges.}
Most seismic stations continuously record ground velocity for several months or years, and operate at a high {\em sampling frequency}, typically from 20 to 200 Hz~\cite{shinohara2021seismometersamplingfrequency}, resulting in massive collections of data series, each containing up to billions of points.
In addition, event localization requires data from multiple stations, and its accuracy increases with the number of sensors, so global seismological archives grow exponentially with the number of deployments~\cite{arrowsmith2022seismologybigdata}. 
This trend is further accelerated by distributed acoustic sensing (DAS), which turns fiber-optic cables into dense seismic arrays with thousands of channels~\cite{zhan2020das}, resulting in massive seismic datasets~\cite{spica2023pubdas}. % datasets already total tens of terabytes~\cite{spica2023pubdas}.
Moreover, seismologists cannot disregard older data, since, ideally, they need to examine the entire seismic record.
Finally, the need to detect multiple types of geophysical events %results in extensive template collections, sometimes 
has led to template collections comprising several tens of thousands of distinct templates.

These factors make efficient template matching highly challenging, due to both the main-memory limitations of a single machine and the extremely large processing times required.
An efficient solution to this challenge would be of considerable importance for seismological observatories and seismic monitoring, enabling the automatic detection of aftershocks and repeating earthquakes, enabling the construction of more complete seismic catalogs while reducing analysts’ workloads, as well as the identification of similar events in large databases, particularly for nuclear test monitoring by the Comprehensive Nuclear-Test-Ban Treaty Organization (CTBTO)~\cite{CTBTO_IDC}.
%However, the growing volume of seismic data, increasingly extensive reference databases, and expanding availability of continuous recordings spanning decades highlight the need for substantial improvements in computational efficiency and search performance.
We stress that in these applications accuracy is of paramount importance, so \emph{exact} answering techniques are a necessity. 
In the following, we focus on exact methods.

Despite important advances in accelerating template matching~\cite{beauce2018fast}, its computational cost remains a major limitation when large datasets are analyzed.
To keep the computation feasible, seismologists often reduce the number of templates considered before running the search, for example by selecting well-located earthquakes as templates or removing redundant events. 
They may also downsample the waveforms, analyze fewer stations and channels, or restrict the explored detection parameters, e.g., the correlation threshold, frequency band, and template length, since each new configuration requires another search over the continuous data.
Decade-long archives of permanent networks and dense deployments with hundreds of sensors are therefore rarely analyzed as a whole, and a catalog is rarely re-scanned when it changes. 
A substantial speedup would thus allow more complete analyses of large seismic datasets, covering broader regions and longer time periods. 
This is all the more important for DAS, where template matching improves earthquake detection~\cite{li2018templatematching} but must scale to thousands of channels.

\noindent \textbf{Our Approach.}
In this paper, we introduce {\em TREMOR}, an efficient distributed solution for large-scale template matching across massive waveform data series collections and large template catalogs.
Our approach operates by conceptually extracting all possible seismic subsequences of a long waveform in parallel, normalizing them on the fly through hardware-conscious optimizations, and creating a distributed data series index across multiple computing nodes.
TREMOR builds on top of Odyssey~\cite{chatzakis2023odyssey}, the state-of-the-art framework for distributed similarity search over full-sequence data series, and employes a data replication technique: machines index the same parts of a waveform, which enables highly efficient query scheduling and manages the fundamental tradeoff between scalability and memory requirements. % with reasonable main-memory requirements.

Driven by the requirements of the applications and analysis pipelines on which the seismology community is focusing, we develop TREMOR to support both top-$k$ and threshold-based template matching, under the ED distance and the Pearson correlation coefficient, which are widely used in seismology.
TREMOR answers every template in one of three ways: with a specialized data series index, with a sequential scan of the waveform that abandons the distance computation of a subsequence as soon as it exceeds the search bound, or with FFT-based distance computation~\cite{mueen2022mass}.
On seismic waveforms, the lower bounds of the index prune well for highly selective templates, i.e., for high correlation thresholds and for templates with close matches; % in the waveform; 
for these, the index is the fastest strategy.
For the other templates, where the index pruning rates are low, scanning the waveform in order, e.g., with a skip-sequential scan~\cite{echihabi2020hydra1,zoumpatianos2016ads,echihabi2022hercules}, is faster, unless the search bound is too loose for early abandoning to pay off, in which case FFTs are faster.
For every template, TREMOR uses a small sample of subsequences to estimate how well the index will prune, and chooses its strategy accordingly.
As a result, TREMOR is much faster than any of the existing template matching methods.

We further deploy TREMOR in two real use cases. 
The first use case, SeiFR, monitors seismicity in metropolitan France by searching years of continuous recordings from stations of the French Atomic Energy Commission (CEA) for matches to cataloged earthquakes~\cite{duverger2021ceaseismic}. 
The current pipeline using conventional tools requires downsampling and independent processing of small data subsets, reducing efficiency and accuracy.
The second use case, Maule, examines aftershocks of the 2010 $M_w$~8.8 earthquake in south-central Chile~\cite{delouis2010slip,madariaga2010central,moreno20102010}, recorded for nearly 10 months by about 156 instruments~\cite{beck2014advancing} and studied extensively~\cite{agurto2012seismic,lange2012aftershock,rietbrock2012aftershock,ryder2012large,hicks2014anatomy,ruiz2014seismological}. 
A recent reanalysis expanded the catalog to approximately 500k events, 400k of which were detected by threshold-based template matching~\cite{flores2026fine}; however, a more detailed analysis comes at a very high computational cost with current tools.
Our deployments demonstrates the significant advantages of using TREMOR in these seismological data analysis pipelines.

\noindent \textbf{Contributions.}
We summarize our contributions as follows:

\noindent $\bullet$
We introduce \textit{TREMOR}, an efficient distributed framework for template matching designed to process large-scale seismic data collections, with novel mechanisms for distributed subsequence indexing, distributed threshold-based similarity search, and online data normalization.

\noindent $\bullet$
We argue that no single search strategy is best for seismic template matching in practice, and design TREMOR's search as an index with scan and FFT fallbacks, adaptively chosen for every template based on the estimated pruning quality of the lower bounds, so that it outperforms searching only the index, a skip-sequential scan, and FFT-based search.
We further introduce a two-pass leaf refinement and, for $k$-NN search, a tight first bound obtained from the index and shared by all nodes; all of them return the same \emph{exact} answers.

\noindent $\bullet$
We demonstrate the applicability of our method by deploying it in two representative real-world template-matching scenarios in seismology, namely, seismicity monitoring in metropolitan France, and identification and analysis of aftershocks of the 2010 $M_w$~8.8 earthquake in south-central Chile.

\noindent $\bullet$
We perform an extensive experimental evaluation on both use cases, comparing our approach with the state of the art:
two distributed versions of MASS~\cite{mueen2022mass}, the FFT-based method popular in seismology, and an optimized skip-sequential scan; TREMOR outperforms them by up to 132$\times$, 18$\times$, and 12$\times$, respectively.

\noindent $\bullet$
We make TREMOR's code publicly available, and also contribute to the community the real datasets of our two use cases, which were not publicly accessible before~\cite{tremorArtifacts}.

%\todo{maybe move this in experiments directly --> move to 4.1}

\section{Preliminaries}
\label{sec:preliminaries}

\begin{figure}
	\centering
	\begin{subfigure} [b]{0.13\textwidth}
		\centering
		\includegraphics[width=\textwidth]{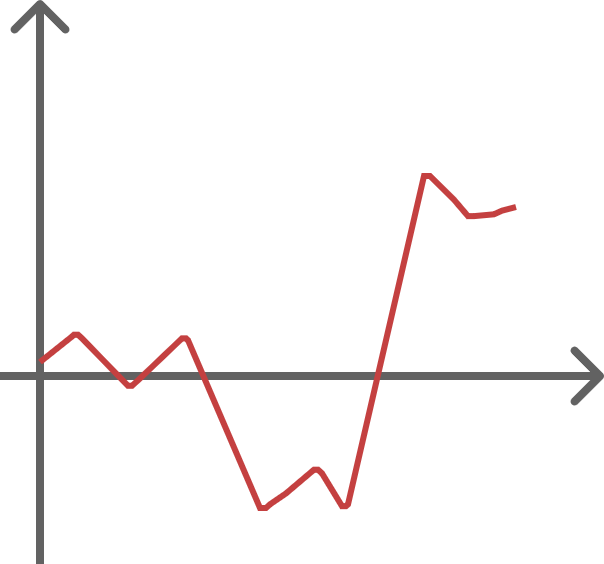}
		\caption{Data Series}
        \label{fig:from_ds_to_iSAX:dataseries}
	\end{subfigure}
	\begin{subfigure} [b]{0.15\textwidth}
		\centering
	\includegraphics[width=0.85\textwidth]{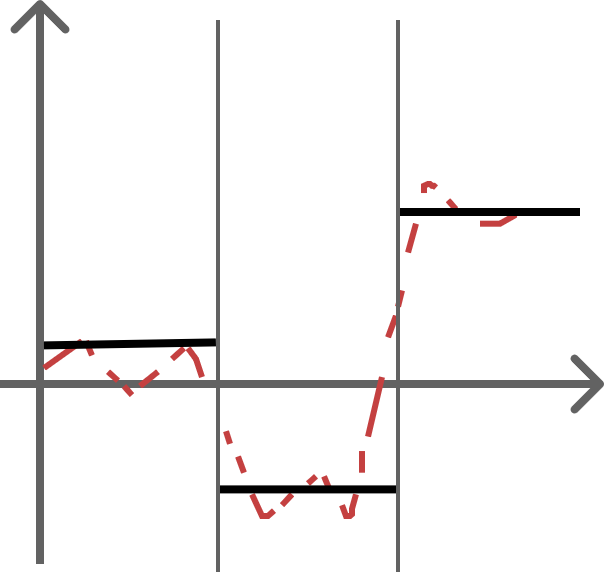}
		\caption{PAA Summary}
		\label{fig:from_ds_to_iSAX:paa}
	\end{subfigure}
	\begin{subfigure} [b]{0.175\textwidth}
		\centering
		\includegraphics[width=\textwidth]{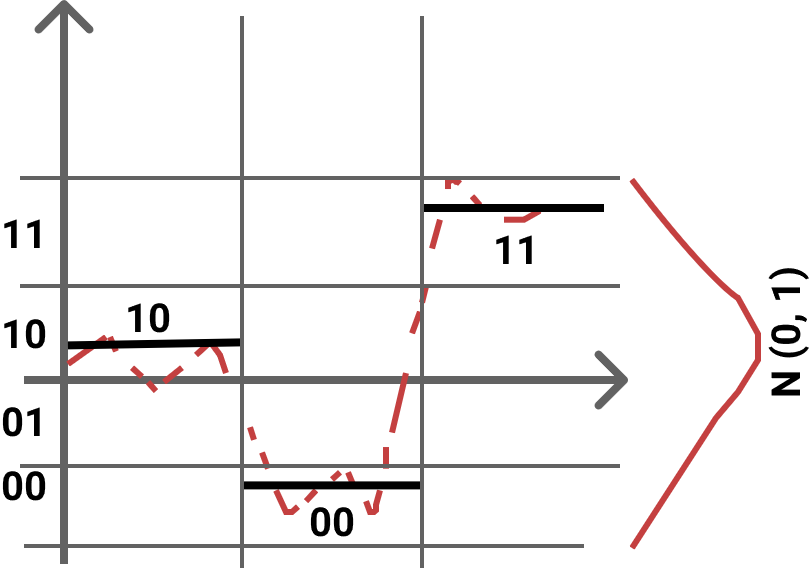}
		\caption{iSAX Summary}
		\label{fig:from_ds_to_iSAX:isax}
	\end{subfigure}

    \vspace{0.2cm}
    
	\begin{subfigure} [b]{0.45\textwidth}
		\centering
	\includegraphics[width=0.9\textwidth]{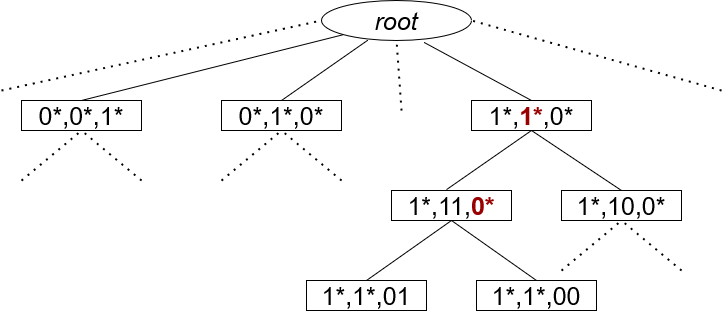}
		\caption{iSAX Tree}
		\label{fig:from_ds_to_iSAX:isaxtree}
	\end{subfigure}
    \vspace{-0.4cm}
	\caption{From data series to iSAX index}
    \vspace{-0.4cm}
	\label{fig:from_ds_to_iSAX}
\end{figure}

%\subsection{Background}
\noindent \textbf{Subsequence Similarity Search (Template Matching).}
A seismic waveform is a data series $\mathit{WF} = {w_1,\dots,w_n}$, where each point $w_i=(u_i,t_i)$ consists of a real-valued seismic activity measurement $u_i$ and its timestamp $t_i$, for $1 \leq i \leq n$. The value $n=|\mathit{WF}|$ is the length, or dimensionality, of the waveform.
Given a waveform $\mathit{WF}$ and a shorter data series, the \emph{query} $Q={q_1,\dots,q_m}$, where $m \ll n$, the goal of \emph{subsequence similarity search} is to find the subsequences of $\mathit{WF}$ of length $m$ that are most similar to $Q$~\cite{echihabi2020hydra1}. 
We denote by $X = \mathit{WF}_{i,m}={w_i,\dots,w_{i+m-1}}$ the subsequence of length $m$ starting at position $i$ of $\mathit{WF}$.
% OLD: In seismology, this task is known as \emph{template matching}, where the query $Q$, i.e., the \emph{template}, represents a seismic event, and $\mathit{WF}$ is a continuous seismic waveform recorded by a seismometer. 
% OLD: Throughout this paper, we use \emph{subsequence similarity search} and \emph{template matching} interchangeably, as well as \emph{query} and \emph{template}.
In seismology, this task is known as \emph{template matching}, where the query $Q$, i.e., the \emph{template}, represents a seismic event, and $\mathit{WF}$ is a continuous waveform. %; we use these terms interchangeably.
%In practice, seismologists search a \emph{collection} of waveforms $\mathit{WFC}={\mathit{WF}_1,\dots,\mathit{WF}_N}$, with one waveform per channel of every station in a seismic network, and each waveform has its own templates, corresponding to events recorded on the same channel. Each template is matched against the waveform of its channel; therefore, template matching over $\mathit{WFC}$ consists of independent subsequence similarity searches, one per waveform. In the following, $\mathit{WF}$ denotes one waveform in the collection.
  
\noindent \textbf{Subsequence Similarity Search Variants.}
% OLD: Subsequence similarity search is broadly categorized into two types.
% OLD: In {\em $k$-nearest-neighbor similarity search ($k$-SS)}, the goal is to find the $k$ closest subsequences, or $k$-NN, of $\mathit{WF}$ to $Q$ based on a similarity or distance metric $M$.
Subsequence similarity search has two variants: in {\em $k$-nearest-neighbor similarity search}, the goal is to find the $k$ closest subsequences, or $k$-NN, of $\mathit{WF}$ to $Q$ based on a similarity or distance metric $M$.
In {\em threshold-based similarity search}, the objective is to retrieve all subsequences of $\mathit{WF}$ whose similarity to $Q$ is above a given threshold $t$, or whose distance is below $t$, depending on $M$.

\noindent \textbf{Distance Measures.}
We focus on two similarity measures that are commonly used in seismology: 
(1) {\em Euclidean distance (ED)}~\cite{kalpakis2001euclideandistance,berthold2016clusteringwitheuclidean,paparrizos2024survey}, defined between the template $Q$ and a subsequence $X = \mathit{WF}_{i,m} = \{x_1, \dots, x_m\}$ as {\small $ED(Q, X) = \sqrt{\sum_{j=1}^{m} (q_j - x_j)^2}$}; and 
(2) {\em cross correlation (CC)}, defined as 
{\small $CC(Q, X) = \frac{\sum_{j=1}^{m} (q_j - \bar{Q})(x_j - \bar{X})}{\sqrt{\sum_{j=1}^{m} (q_j - \bar{Q})^2} \sqrt{\sum_{j=1}^{m} (x_j - \bar{X})^2}}$}, where $\bar{Q}$ and $\bar{X}$ are the mean values of $Q$ and $X$, respectively.
% OLD: Classic template-matching use cases perform {\em z-normalization} on data series, i.e., transforming them into series with mean $\mu=0$ and standard deviation $\sigma=1$~\cite{lima2023znorm,rakthanmanon2012trillionsubsequences,rakthanmanon2013bigdatatimeseries}. 
% OLD: This ensures that similarity search is invariant to amplitude differences.
Classic template-matching use cases perform {\em z-normalization} on data series, i.e., transforming them into series with mean $\mu=0$ and standard deviation $\sigma=1$, which makes the search invariant to amplitude differences~\cite{lima2023znorm,rakthanmanon2012trillionsubsequences,rakthanmanon2013bigdatatimeseries}.

\noindent \textbf{iSAX Summarization and Indexing.}
The {\em iSAX} summary~\cite{camerra2010isax} of a data series partitions the x-axis into equal-length segments and represents each segment using the mean value of the data points it contains, as shown in Figure~\ref{fig:from_ds_to_iSAX}(a--c).
It then divides the y-axis into regions, 
and represents each region using a predefined number of bits, referred to as {\em cardinality}.
The number of bits can vary across regions, enabling the construction of a hierarchical index tree, namely an iSAX-based index tree~\cite{camerra2010isax}, as shown in Figure~\ref{fig:from_ds_to_iSAX}(d).

iSAX-based indexes~\cite{parisplus,peng2020messi,chatzakis2023odyssey,wang2023dumpy,peng2021sing,wang2024dumpyos,fresh} exploit multiple threads and {\em SIMD} to construct an index tree and efficiently answer queries. % using this tree.
They compute the summarizations of all data series in an almost embarrassingly parallel manner, avoiding synchronization overhead and improving locality in data access during tree construction~\cite{messiJournal,fresh}.
To answer a query, they traverse the tree to the most relevant data series, whose distance to the query becomes the initial best-so-far (BSF) answer, and prune every data series whose lower-bound distance to the query, derived from the iSAX summaries, exceeds the BSF.
Multiple threads then compute the real distances of the remaining series, which are maintained in priority queues~\cite{peng2020messi,chatzakis2023odyssey}, updating the BSF if a better answer is found.

\subsection{Related Work}
\label{sec:related-work}

\noindent \textbf{Data Series Similarity Search.}
%In recent years, 
The data management community has proposed several efficient solutions for in-memory data series and vector similarity search~\cite{peng2020messi,peng2021sing,chatzakis2023odyssey,chatzakis2026odysseyplus,wang2023dumpy,wang2024dumpyos,fresh,palpanas2020dsindexevolution,del2026daisy,malkov2018hnsw, chatzakis2025darth, chatzakis2026vader, wang2026annie}. 
%\tp{edw 9a mas thn poyne poy den leme tipota gia ta subsequence similarity search indexes/solutions}
% OLD: However, these solutions are primarily designed for \emph{full-sequence} similarity search, where the query $Q$ and the data series of the collection have the same length.
% OLD: ... where both the query $\mathcal{S}$ and data sequence $\mathcal{C}$ have the same dimensionality.
% OLD: In this work, we focus on subsequence similarity search, where every waveform $\mathit{WF}$ of the collection is much longer than the template $Q$ (Section~\ref{sec:preliminaries}).
However, these solutions are primarily designed for \emph{full-sequence} similarity search (or whole-matching), where the query $Q$ and the series in the collection have the same length, whereas template matching is a subsequence similarity search problem~\cite{echihabi2020hydra1}.
% OLD: In this work, we focus on a different problem, where the dimensionality of $\mathcal{C}$ is much larger than that of $\mathcal{S}$. \tp{check again after fixing sec2.1}

\noindent \textbf{Distributed Data Series Processing.}
% OLD: Since modern data series processing tasks involve very large data collections that may require extensive processing times or may not fit entirely in the memory of a single machine, distributed frameworks are essential.
% OLD: Odyssey~\cite{chatzakis2023odyssey, chatzakis2026odysseyplus} is the state-of-the-art framework for in-memory data series similarity search, leveraging the power of modern clusters composed of multiple computation nodes.
% OLD: Based on the hardware specifications of the underlying infrastructure, Odyssey employs an adaptive data replication scheme to distribute data across available nodes. %, optimizing for the nature of the task.
% OLD: Additionally, it supports various load-balancing techniques, including query scheduling based on execution-time predictions, work-stealing, and BSF sharing.
Odyssey~\cite{chatzakis2023odyssey, chatzakis2026odysseyplus} is the state-of-the-art distributed framework for in-memory data series similarity search: it distributes the data across nodes using adaptive replication, and balances the load with query scheduling based on execution-time predictions, work-stealing, and BSF sharing.

\noindent \textbf{Seismic Template Matching Approaches.}
Template matching in seismology has been implemented using different approaches for data series subsequence similarity search~\cite{shabikay2024matrix,mueen2022mass,rakthanmanon2012trillionsubsequences,linardi2018ulisse,twinsubsequences,beauce2024bpmf}.
The state-of-the-art approach is the widely used {\em MASS} algorithm~\cite{mueen2022mass}.
MASS efficiently computes the z-normalized Euclidean distance between a query and all waveform subsequences of a long data series using the {\em fast Fourier transform (FFT)}~\cite{cooley1965fft,frigo1998fftw}.
% OLD: Instead of performing a brute-force comparison, MASS leverages FFT to transform the problem into the frequency domain, where calculations are significantly faster.
Additionally, the {\em ULISSE} algorithm~\cite{linardi2018ulisse,linardi2020ulisse} %addresses the challenge of performing subsequence similarity search on data series of varying lengths.
%It introduces a novel summarization technique that succinctly represents multiple sequences of different lengths within a single index structure.
%This approach enables 
was proposed to enable efficient approximate and exact similarity search with variable-length queries.
% \new{
% We do not compare with ULISSE, since it targets queries of varying lengths on a single machine; for templates of a single length, as in template matching, its index is an iSAX index over the subsequences, like the one that TREMOR distributes, and Section~\ref{sec:eval:strategies} compares this index with scans.
% }
% OLD: (after "...of varying lengths.") \tp{why do we not compare to this?}
% OLD: Another solution is the fast matched-filter approach (FAST)~\cite{beauce2018fast}, which computes the correlation coefficients of z-normalized data series and is implemented in CUDA for GPU support, following a similar approach to MASS.
% OLD: Although similarity search is fast, this approach exhaustively searches for similar matches without further optimization.
Another solution is the fast matched-filter approach (FAST)~\cite{beauce2018fast}, which computes the correlation coefficients of z-normalized data series on GPUs, similarly to MASS, and exhaustively searches for matches without further optimization.

%\vspace{-0.51cm}
\section{The TREMOR Framework}
\label{sec:tremor}

\begin{figure*}[t]
    \centering
    \includegraphics[page=1,width=0.95\textwidth]{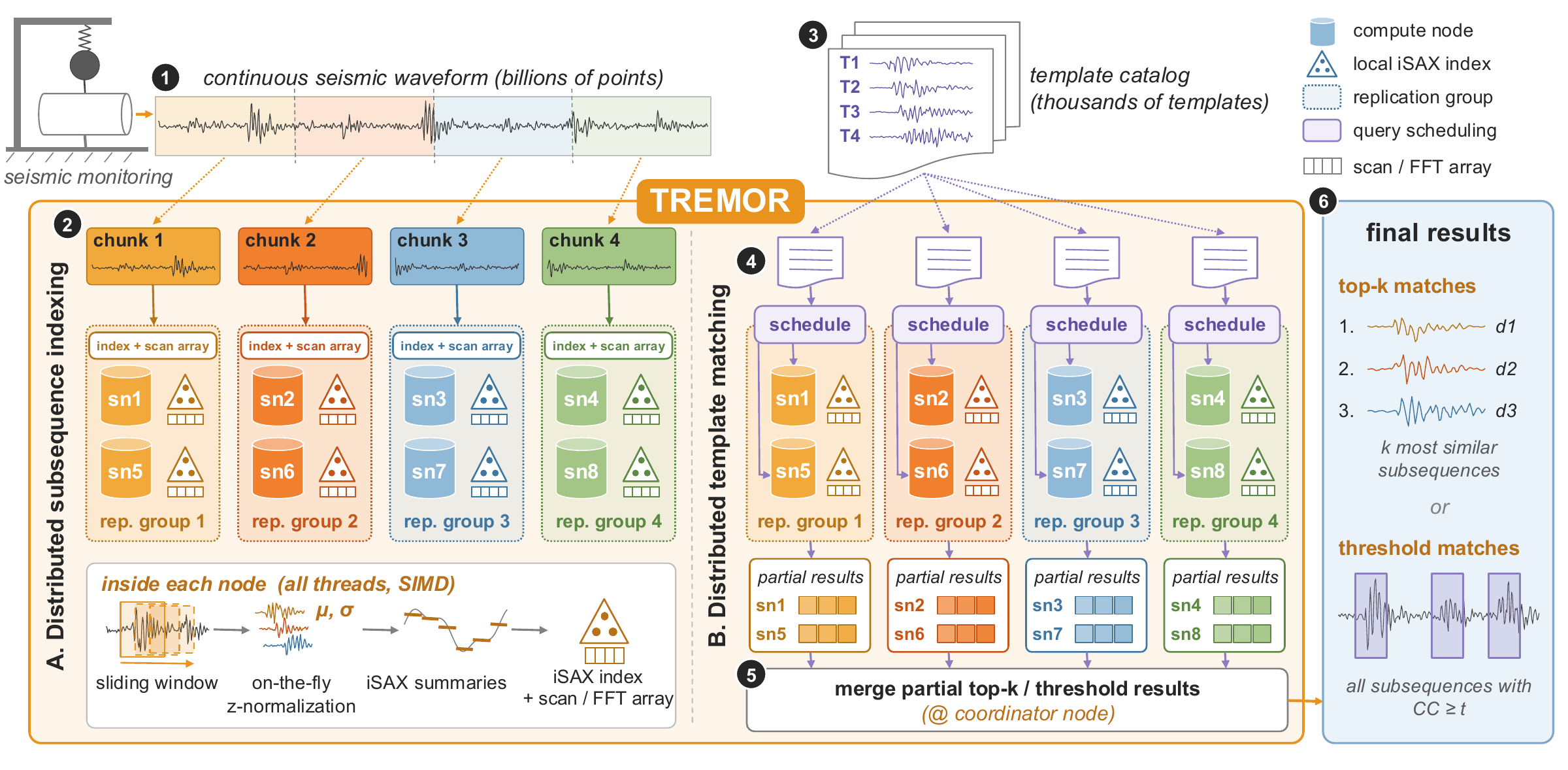}
    \vspace{-0.6cm}
    \caption{Overview of TREMOR.}
    \vspace*{-0.3cm}
    \label{fig:overview}
\end{figure*}

An overview of TREMOR is presented in Figure~\ref{fig:overview}.
TREMOR operates on top of a long seismic waveform: it partitions the waveform into a tunable number of chunks and then assigns each chunk to a group of nodes for indexing.
Note that (due to replication) more than one node may index the same part of a waveform, allowing TREMOR to perform efficient query scheduling~\cite{chatzakis2023odyssey}.
The group of nodes that index the same partition of the waveform is called a replication group.
In the example shown in the figure, the system consists of 4 replication groups (partial-4) and 8 nodes in total, but both the number of replication groups and their capacity are fully tunable in our system.
Each node then runs a sliding window over its assigned part of the waveform, extracting subsequences of a predefined length and normalizing them on the fly.
It then constructs a local iSAX index that stores the z-normalized subsequences extracted from the corresponding part of the waveform.

Once all nodes in the system finalize index construction, which can typically be completed in just a few minutes even for waveforms containing billions of samples, TREMOR is ready to process templates.
The template catalog is provided in full to TREMOR, and TREMOR replicates the catalog to each replication group.
Then, in each replication group, each template is assigned to only one node of the replication group, since these nodes hold the same data and only one node is sufficient to process the template within each replication group.
This assignment is performed using a dynamic scheduling policy~\cite{chatzakis2023odyssey}, where nodes are assigned one template at the beginning and receive a new one only after they finish their current work, resulting in a highly effective load-balancing scheme.
At the same time, %building on prior work, 
TREMOR nodes can %optionally 
share their results across the system whenever a closer sequence to a template is found by broadcasting their BSF, ensuring that nodes in other replication groups process templates with the global best results, which leads to better pruning and even better load balancing across the system.
% OLD: Note that TREMOR can process the given template catalog using either top-$k$ search or threshold-based search.
Once all templates in the catalog have been processed, the results are gathered in a single coordinator node, which performs the merge, and returns the final results.

% OLD: For the rest of this section, we discuss the details of the mechanisms provided by TREMOR, with special focus on the features we developed to support efficient template matching.

\subsection{Distributed Subsequence Indexing}
Although data replication and distributed iSAX indexing have been established in Odyssey~\cite{chatzakis2023odyssey}, Odyssey does not support subsequence indexing; TREMOR is the first framework to support distributed iSAX-based subsequence indexing for template matching over long waveforms.

To index subsequences, each TREMOR node runs a sliding window over its waveform chunk, and conceptually extracts all subsequences one by one (keeping only a pointer to their starting position).
TREMOR z-normalizes every subsequence, which requires its mean and standard deviation; since recomputing them for billions of subsequences would lead to significant indexing overheads, TREMOR computes them only for the first subsequence, and then updates them for each new subsequence based on the added and omitted values.
It then inserts the iSAX summary of every z-normalized subsequence into a node-local iSAX index, keeping pointers to the raw waveform instead of the subsequences themselves to keep memory requirements to a minimum, and caches the mean and standard deviation of every subsequence for the distance computations.
TREMOR also keeps the iSAX summaries in waveform order, to easily switch between index search and the sequential scan (Section~\ref{sec:tremor:adaptive}).
%}

\subsection{Distributed Template Matching}
Once each node in the system has built its local iSAX index, TREMOR is ready to process the catalog of templates, using the mechanisms described in this section. 

In template matching, all templates have the same length, which also defines the size of the extracted subsequences. 
Depending on the seismic pipeline, the value $k$ for top-$k$ similarity search, or the threshold for threshold-based similarity search is provided.
TREMOR  (in contrast to Odyssey) supports both the Euclidean distance (ED) and cross-correlation (CC) similarity measures.

\subsubsection{Query Scheduling}
To process the templates, TREMOR estimates the execution time of each template and reorders the templates so that those estimated to require more processing time are processed first.
Then, within each replication group, a designated replication-group coordinator dynamically assigns templates to nodes for processing whenever a node finishes processing its previous query.
This dynamic scheduling technique, especially when templates are sorted so that the most computationally intensive ones are processed first, is highly effective~\cite{chatzakis2023odyssey}.

\subsubsection{Load Balancing Mechanisms}
TREMOR also builds upon load-balancing mechanisms of Odyssey~\cite{chatzakis2023odyssey,chatzakis2026odysseyplus} for cases in which the templates in the catalog may require very different processing times and lead to disproportionate workloads across nodes.
TREMOR implements BSF sharing across the entire system, meaning that whenever a node finds an improved distance or similarity between a subsequence and a template, it shares this value with all nodes in the system.
This allows nodes in other replication groups that may be processing, or will later process, the same template to perform the search with a better BSF value, thus improving pruning and reducing search times.
This optimization ensures load balancing across nodes in different replication groups that process the same templates.
Furthermore, since imbalances may arise even among nodes in the same replication group, e.g., when one node is assigned a difficult template that requires a long processing time while the other nodes finish early and remain idle, TREMOR implements work stealing among nodes in the same replication group.
Work stealing automatically detects nodes that are still processing their assigned templates after other nodes have finished their own work and transfers part of this work to idle nodes, enabling faster processing and avoiding idle time while other nodes are still working.
For more details on these mechanisms and their implementation, we refer interested readers to~\cite{chatzakis2023odyssey}.

\subsubsection{Local Node Template Matching}
\label{sec:LocalNodeTemplateMatching}
When a template is scheduled for processing at a node in the system, the local iSAX index is searched to locate the subsequences that are most similar to the given template.
Each node initializes a priority queue to collect the qualifying subsequences.
In the case of top-$k$ similarity search, this priority queue has an upper size bound of $k$, while threshold search uses the same priority queue without bounds.
The search follows the same procedure as classic iSAX indexing, described in Section~\ref{sec:preliminaries}, with the BSF answer used for pruning being either the $k$-th distance in the priority queue or the threshold value, depending on the search variant.
In addition, the search can be conducted using either ED distance or CC, which are similar for z-normalized data.
Recall that, for optimization reasons, TREMOR does not store the actual raw extracted subsequences in the leaves of the index, but only stores start and end pointers to the original waveform, which is kept as a simple raw data array in memory.
For this reason, once the search for a template reaches an index leaf to calculate the actual distances, rather than the lower-bound ones, the corresponding subsequence must be z-normalized again.
In TREMOR, this is a highly optimized procedure because the mean and standard deviation of each subsequence, which are needed for z-normalization, are stored in the index (and thus not recomputed).
Moreover, the z-normalization procedure is combined with the raw distance calculation between the subsequence and the template, enabling 
efficient search even when real distances are required.

\noindent \textbf{Early-Abandoning Distance Computation.}
TREMOR computes the squared distance between a template and a candidate subsequence with SIMD instructions and abandons it as soon as the partial sum exceeds the current bound, i.e., the threshold distance or the $k$-th BSF distance.
Following the UCR suite~\cite{rakthanmanon2012trillionsubsequences}, the template is split into blocks of 32 points that are visited in decreasing order of the template's energy, so that the blocks contributing most to the distance of a non-matching subsequence are accumulated first; the bound is checked once per block.

\noindent \textbf{Two-Pass Leaf Refinement.}
When the search reaches a leaf, TREMOR first computes the lower bounds of all subsequences in the leaf (using the iSAX summaries), and then computes the distances of the subsequences that pass, prefetching the statistics and the first visited blocks of upcoming ones to hide part of the memory latency caused by their scattered positions in the waveform.

\noindent \textbf{Adaptive Search: Index with Scan Fallback.}
\label{sec:tremor:adaptive}
The index helps only when its lower bounds prune most subsequences.
Seismic waveforms oscillate around zero, so after z-normalization the segment means that iSAX summarizes are close to zero for several subsequences, and the lower bounds are too small to prune, except when the search bound is very tight: for high correlation thresholds, and for $k$-NN search of templates with close matches.
When the index leaves almost every subsequence as a candidate, checking the candidates leaf by leaf jumps randomly across the waveform; it is then much faster to scan the subsequences in their order in the waveform, since consecutive subsequences overlap and their points are already in the cache.
TREMOR therefore combines its index with a scan fallback, and decides how to answer every template from the lower bounds of an evenly spaced sample of $2^{16}$ subsequences per node, i.e., from the share $c$ of sampled subsequences whose lower bound is within the search bound:

\noindent(1) 
if $c \leq f$ (1\% by default in all our experiments), the lower bounds prune almost every subsequence, and TREMOR searches the index;
    
\noindent(2) 
if $f < c < b$ (5\% by default in all our experiments), TREMOR falls back to a \emph{skip-sequential scan}~\cite{zoumpatianos2016ads, echihabi2020hydra1, echihabi2022hercules}: it visits the subsequences in waveform order, reads the iSAX summary of each (kept in waveform order next to the tree), skips the subsequence when its lower bound exceeds the search lower bound, and otherwise computes its real distance;
    
\noindent(3) 
if $c \geq b$, TREMOR falls back to an early-abandoning sequential scan without lower bounds, or to FFT-based distance computation. 

The third strategy matters because the scan is bound by memory bandwidth: reading the iSAX summary of a subsequence and computing its lower bound costs about as much as abandoning its distance after the first block, so the lower bounds pay off only when they prune almost every subsequence.
In the leaves, where a subsequence whose lower bound is not checked costs a distance computation at a scattered position, TREMOR skips the lower bounds only for threshold search, when at least half of the sampled subsequences pass; for $k$-NN search, the bound tightens during the search, so the leaves keep them.

\noindent \textbf{FFT Fallback.}
\label{sec:tremor:fft}
Early abandoning pays off only when the search bound is tight. 
When it is loose, e.g., for low correlation thresholds, or for $k$-NN search whose $k$-th answer is far from the template, the scan examines most points of every subsequence.
TREMOR then computes the dot products of the template with all subsequences with FFTs, similar to MASS~\cite{mueen2022mass}, over pieces of $2^{15}$ points that overlap by $m-1$ points and whose FFTs are stored in the index, and uses the resulting distances only as a filter: every subsequence within the search bound, up to the FFT rounding error, is verified by computing its exact distance to the query, ensuring that the answers obtained from FFT are exact.
TREMOR chooses between the scan and the FFTs for every template, by estimating the cost of each alternative.
The scan checks a subsequence in blocks of 32 points and stops as soon as the partial distance exceeds the search bound, so its time grows with the number of blocks that it visits per subsequence.
While building its index, every node measures (on its data) the time of an FFT pass, which is the same for every template, since it does not depend on the threshold or $k$ value, and the time of the scan per subsequence when it stops after the first block, and for each additional block.
For every template, the node runs the scan with the current search bound on a sample of subsequences, and predicts the time of the full scan from the average number of blocks visited.
It uses the FFTs when they are predicted to be at least 10\% faster than the scan; this happens when the bound is loose and the scan would examine most blocks of every subsequence.

\noindent \textbf{First Bound for $k$-NN Search.}
A $k$-NN search can prune only once it has found candidate matches that bound the answer.
Before the search, TREMOR computes (using all node threads) the distances of the subsequences in the leaf that the template's own iSAX word falls into, i.e., the classic approximate iSAX search~\cite{camerra2010isax}.
For a template with a close match in the waveform, this yields a tight bound from the start, which lets TREMOR use the index and makes early abandoning effective in the scan.
Without replication, the close match of a template lies in the part of the waveform of a single node, so the nodes exchange their candidate matches before the search, and every node prunes with the overall best first bound. % found anywhere.

\noindent \textbf{Role of the Index.}
The index alone answers only selective templates, i.e., for high correlation thresholds and small values of $k$, while the others are answered with a scan or with FFTs.
Note that in all these cases the index is necessary; it effectively takes on the roles of cost estimation (based on amount of pruning) and access path selection. 
TREMOR to choose adaptively among all the available strategies for every template as follows:
its iSAX summaries provide the lower bounds from which TREMOR estimates the portion $c$ of subsequences that survive pruning, and, for $k$-NN search, its approximate search provides the first search bound, without which neither the pruning of the lower bounds nor the cost of the early-abandoning scan, which TREMOR compares to that of the FFTs, can be estimated.
Hence, the index is the mechanism that lets TREMOR select the most efficient strategy for every template, and thus outperform each strategy alone.
%\noindent \textbf{Per-Template Overheads.}
%Selective templates take a few milliseconds per node, so per-template overheads matter: TREMOR keeps its query threads alive and pinned to cores across templates, and reuses its per-template buffers instead of allocating them for every template.

\noindent \textbf{Merging Nearby Matches.}
A seismic event matches not only one subsequence but also its neighbors, which overlap it.
To report each event once, TREMOR merges matches that are closer than a \emph{merge offset} (a quarter of the template length in our experiments).
In threshold search, it keeps the first match of each group of nearby matches.
In $k$-NN search, it returns the $k$ best matches that are at least the merge offset apart: it takes the matches from best to worst and skips any match that is too close to one already taken.
To find these matches without checking every subsequence, TREMOR prunes with a slightly looser bound than in plain $k$-NN search: it keeps $2k-1$ matches that are pairwise at least the merge offset apart, and prunes with the distance of the worst of them.
Each of the first $k-1$ answers can exclude at most two of these matches, so one of them is still eligible as the $k$-th answer, and no answer is missed; for $k=1$, the bound is the nearest-neighbor distance.
Each node keeps the matches within this bound, and the coordinator selects the final $k$ matches.

\section{Experimental Evaluation}
\label{sec:evaluation}

\noindent \textbf{Setup.}
We use up to 8 nodes, each with two AMD EPYC 7H12 processors (128 cores at 2.6~GHz) and 256~GB of RAM, connected through HDR 100 InfiniBand network.
All methods implemented in C/C++ with MPI. 
Our code and datasets are publicly available~\cite{tremorArtifacts}.

\noindent \textbf{Algorithms.}
We compare our approach with MASS~\cite{mueen2022mass}, the currently popular approach for seismic template matching.
The publicly available C/C++ version of MASS, built on the FFTW library, is neither parallel nor distributed, and supports only top-1 search.
We therefore implement DMASS, which we also make publicly available~\cite{tremorArtifacts}: it uses the multithreaded FFTs of FFTW and the same replication strategies as TREMOR, and it selects the $k$-NN and threshold answers with the same priority queue and exclusion-zone rule as TREMOR, so that both return the same answers.
We evaluate two versions of DMASS: \emph{DMASS-V1}, as the original MASS, computes one FFT over each node's whole part of the waveform, whereas \emph{DMASS-V3}, as MASS V3~\cite{mueen2022mass}, computes the FFTs over overlapping pieces of $2^{15}$ points, which the threads of the node process in parallel.
We also compare with a \emph{skip-sequential scan} (SSS), the standard index-free baseline of data series similarity search~\cite{echihabi2020hydra1, zoumpatianos2016ads, echihabi2022hercules}.
SSS keeps the iSAX summary, mean, and standard deviation of every subsequence in waveform order, and answers a template by scanning the subsequences in order: it skips a subsequence when its lower bound exceeds the search bound, and otherwise computes its distance with the same early-abandoning kernel as TREMOR (Section~\ref{sec:tremor}).
SSS uses the same partitioning of the waveform, replication strategies, and dynamic template scheduling as TREMOR.

\noindent \textbf{TREMOR Configuration.}
TREMOR combines its index with the scan fallback of Section~\ref{sec:tremor:adaptive}, with $f=1\%$ and $b=5\%$, two-pass leaf refinement, and the first $k$-NN bound from the index, and with the FFT fallback of Section~\ref{sec:tremor:fft}.
Its index has leaves of 2K subsequences, and the number of PAA segments is chosen per template length (16 for 496- and 992-point templates, 10 for 1K-point templates, and 8 for 3K-point templates; Section~\ref{sec:eval:parameters}).

\noindent \textbf{Performance Measures.}
Our main performance measure is the \emph{query answering time}: the time from the moment the templates are available until all results are written, on the slowest node.
TREMOR's index construction and DMASS's preprocessing (data loading, FFT, and subsequence statistics) are performed once per waveform and template length and are reused by every template, so we exclude them from the comparison and study TREMOR's index construction separately (Section~\ref{sec:eval:ingestion}).
%Unless otherwise stated, we use 4 nodes with full replication: every node holds the whole waveform, and each template is answered by one node; Section~\ref{sec:eval:replication} studies the other strategies.
All methods answer the same templates of every channel (Section~\ref{sec:eval:workloads}).
We report the total query answering time over the channels of a dataset, and the \emph{speedup} of a method over another.
All methods are exact, and we verify that they return the same answers in every run.

\subsection{Use Cases: Datasets and Workloads}
%\todo{We have deployed TREMOR into two use cases ...... at IPGP and CEA and we use two real use cases, one from each site in this study. }
We deploy TREMOR in two real-world use cases, at two different sites in France, namely, the Institut de Physique du Globe de Paris (IPGP), and the Commissariat à l'Energie Atomique (CEA). 
We deescribe the two uses cases below.

\begin{table}[t]
    \centering
    \scriptsize
    \begin{tabular}{l|rr}
        \toprule
         & SeiFR               & Maule                      \\
        \midrule
        Channels               & 7             & 243                  \\
        Sampling rate                & 25~Hz               & 50 or 100~Hz               \\
        Points per channel     & 1.05--1.58 billion & 8.6 million--1.82 billion \\
        Points (size)          & 10.5 billion (42~GB) & 166.5 billion (665~GB) \\
        Templates             & 4,430         & 11,922              \\
        Template length (points)     & 3,000 (120s)       & 496, 992, 1,000 (10s)     \\
        Search                & $k$-NN, $k \in \{1, 2, 5, 10\}$ & threshold, $t \in [0.6, 0.999]$ \\
        \bottomrule
    \end{tabular}
    \captionsetup{position=bottom}  % the caption is below the table
    \vspace{-0.2cm}
    \caption{Datasets used in the evaluation\todo{}}
    \label{tab:eval:datasets}
    \vspace{-0.4cm}
\end{table}

\subsubsection{Use Case 1: Maule}
\label{sec:eval:maule}
Maule concerns the aftershocks of one of the largest earthquakes ever recorded, the $M_w$~8.8 Maule earthquake that struck south-central Chile in 2010~\cite{delouis2010slip,madariaga2010central,moreno20102010}, recorded for almost 10 months by a dense temporary seismic network of about 156 instruments months~\cite{beck2014advancing}.
These records have motivated many studies~\cite{agurto2012seismic,lange2012aftershock,rietbrock2012aftershock,ryder2012large,hicks2014anatomy,ruiz2014seismological}, and their recent reanalysis with deep-learning-based algorithms to detect earthquakes and threshold-based template matching increased the initial catalogs up to approximately 500K events, from which 400K corresponds to template-matching~\cite{flores2026fine}.
Maule consists of continuous recordings of the International Maule Aftershock Deployment (IMAD), a dense temporary network deployed for 10 months after the magnitude-8.8 earthquake that struck Maule, Chile, in 2010.
We use the 243 channels of this deployment; Table~\ref{tab:eval:datasets} summarizes their characteristics.
The templates of a channel are 10-s recordings (496, 992, or 1,000 points, depending on the channel and sampling rate) of earthquakes detected in this deployment~\cite{flores2026fine}: from 3 to 20K templates per channel, and 1.3 million in total.
In the reference Maule analysis, matched filtering scans the continuous waveform data and computes the cross-correlation across the available channels. 
%A detection is retained when the network-averaged correlation exceeds a time-dependent threshold defined as 8 times the RMS of each 30-min correlation segment, using at least 3 stations and 6 channels and up to 10 stations per template~\cite{beauce2018fast,beauce2024bpmf,flores2026fine}.
%The waveforms and templates are preprocessed as in SeiFR.
Gaps are filled with Gaussian noise, and the waveforms and templates are detrended, tapered, and band-pass filtered between 1-10~Hz.

\subsubsection{Use Case 2: SeiFR}
\label{sec:eval:seifr}
SeiFR concerns the monitoring of the seismicity of metropolitan France. 
Seismologists are interested in searching years of continuous recordings of multiple stations of CEA for the closest matches (i.e., $k$-NN search) of template earthquakes included in the CEA catalog~\cite{duverger2021ceaseismic}. 
The current analysis pipeline includes both down-sampling and splitting the dataset to small (easy to handle) subsets that need to be processed independently, making this analysis time-consuming, cumbersome, and less accurate, 
%\new{
an issue which we solve with the distributed and node-specific mechanisms of (challenges that are addressed by TREMOR).
%}
%\tp{ki emeis pws to kanoyme twra?}
SeiFR consists of up to two years (January 2020 to December 2021) of continuous recordings of seven broadband stations.
The templates of a channel are 120s recordings (3,000 points) of the earthquakes, each starting 2s before an arrival picked on the channel: from 279 to 873 templates per channel, and 4430 in total.
Data preprocessing is done in a similar way as Maule.

\subsubsection{Tasks}
\label{sec:eval:workloads}
Table~\ref{tab:eval:datasets} summarizes the datasets.
In both of our deployments, the goal of seismologists is to perform template matching using predefined event catalogs.
In Maule, we are specifically interested in threshold search, while in SeiFR, we target $k$-NN search.
Following the practice of each use case, we run threshold search on Maule, with correlation thresholds $t \in \{0.6, 0.7, 0.8, \allowbreak 0.9, \allowbreak 0.95, \allowbreak 0.99, \allowbreak 0.999\}$, and $k$-NN search on SeiFR, with $k \in \{1, 2, 5, 10\}$.
On SeiFR, all methods answer all 4,430 templates; on Maule, they answer the same 50 templates of every channel, sampled uniformly at random with a fixed seed (all templates when a channel has fewer), i.e., 11,922 templates.
In both cases, the merge offset is a quarter of the template length (cf. Section~\ref{sec:tremor}).

\subsection{Results}

%\tp{any results with varying dataset size and template size?}

\label{sec:eval:results}

% \begin{figure}[t]
%     \centering
%     \includegraphics{figs/results/legend_methods.pdf}\\
    
%     \begin{minipage}[b]{0.24\textwidth}
%         \centering
%         \includegraphics[width=\linewidth]{figs/results/query_time_vs_threshold_maule_4nodes_repl-full.pdf}
%         \subcaption{Maule}
%     \end{minipage}
%     \begin{minipage}[b]{0.24\textwidth}
%         \centering
%         \includegraphics[width=\linewidth]{figs/results/query_time_vs_k_seifr_4nodes_repl-full.pdf}
%         \subcaption{SeiFR}
%     \end{minipage}
    
%     \caption{Query answering time (log scale) with full replication on 4 nodes as a function of the correlation threshold $t$ (Maule) and of $k$ (SeiFR)}
%     \label{fig:eval:params}
% \end{figure}

% \begin{figure}[t]
%     \centering
%     \includegraphics{figs/results/legend_methods.pdf}\\
%     \begin{minipage}[b]{0.24\textwidth}
%         \centering
%         \includegraphics[width=\linewidth]{figs/results/query_time_vs_nodes_maule_th0.9_repl-full.pdf}
%         \subcaption{Maule, varying nodes ($t=0.9$)}
%     \end{minipage}
%     \hfill
%     \begin{minipage}[b]{0.24\textwidth}
%         \centering
%         \includegraphics[width=\linewidth]{figs/results/query_time_vs_nodes_seifr_k5_repl-full.pdf}
%         \subcaption{SeiFR, varying nodes ($k=5$)}
%     \end{minipage}
%     \caption{Query answering time (log scale) with full replication, (a, b) on 4 nodes as a function of the correlation threshold $t$ (Maule) and of $k$ (SeiFR), and (c, d) as a function of the number of nodes.}
%     \label{fig:eval:nodes}
% \end{figure}

\subsubsection{Varying the Threshold and $k$}
\label{sec:eval:competitors}
Figure~\ref{fig:eval:params} presents the query answering time of the four methods for varying threshold on Maule and varying $k$ on SeiFR, on 4 nodes under full replication (we examine other replication degrees in Section~\ref{sec:eval:replication}).
TREMOR is the fastest method on every workload.
On Maule, it is 2.3--2.9$\times$ faster than SSS, 9.3--18$\times$ faster than DMASS-V3, and 69--132$\times$ faster than DMASS-V1, and on SeiFR, 5.3--12$\times$, 6.5--7.9$\times$, and 56--69$\times$ faster.
The baselines lose for different reasons.
SSS relies on lower bounds and early abandoning: the lower bounds prune little for lower thresholds, and early abandoning stops paying off when the search bound is loose.
Its time on Maule therefore grows as the threshold decreases, and on SeiFR, it is slower for $k \geq 2$, where the $k$-th answers are far from the templates and the scan examines a large fraction of the $m=3K$ points of every subsequence; there, SSS is slower than DMASS-V3.
Both versions of DMASS compute the correlation of the template with every subsequence of the waveform independently of the threshold, thus, their performance remains roughly stable across varying values of threshold and k.
TREMOR chooses its strategy adaptively for every template: it answers selective templates through its index, scans with early abandoning when the search bound is tight enough, and uses FFTs otherwise.

\subsubsection{Varying the Number of Nodes}
\label{sec:eval:scalability}

Figure~\ref{fig:eval:nodes} shows the query answering time on 1-8 nodes under full replication ($t$=0.9 on Maule; $k$=5 on SeiFR).
TREMOR is the fastest method at every node count: on Maule, it is 2.2--2.9$\times$ faster than SSS, 11--14$\times$ faster than DMASS-V3, and 99--115$\times$ faster than DMASS-V1, and on SeiFR, 12$\times$, 5.7--6.7$\times$, and 56--61$\times$ faster.
All methods benefit from more nodes: from 1 to 8 nodes, TREMOR's query time drops from 1,273 to 246s on Maule (5.2$\times$) and from 1,728 to 223s on SeiFR (7.8$\times$); SSS becomes 6.9--7.9$\times$ faster, DMASS-V3 6.4--7.1$\times$, and DMASS-V1 6.0--7.3$\times$.

\subsubsection{Varying the Replication Degree}
\label{sec:eval:replication}
Figure~\ref{fig:eval:repl} compares the methods under varying replication degrees on 4 nodes.
Without replication, each node searches a different part of the waveform for every template, and the query time is that of the slowest node for each template.
Full replication is the fastest configuration of TREMOR on both datasets: it is 1.7$\times$ faster than no replication on Maule (377 vs.\ 633s) and 1.1$\times$ faster on SeiFR (447 vs.\ 483s).
TREMOR is the fastest method at every replication degree: it is 1.7--12$\times$ faster than SSS, 6.0--14$\times$ faster than DMASS-V3, and 56--108$\times$ faster than DMASS-V1.
The baselines gain little from replication: full replication makes them at most 1.4$\times$ faster than no replication.
TREMOR exploits replication through its query scheduling 
(cf. Section~\ref{sec:tremor}): with replication, every template is answered by a single node, and the nodes balance their load dynamically.

\subsubsection{Varying Dataset Size}
% OLD: Figure~\ref{fig:eval:dssize_ingestion}(a) reports TREMOR's query answering time on 4 nodes as the dataset size increases (as a percentage of the points of each dataset), for Maule with $th=0.9$ and SeiFR with $k=5$.
% OLD: Due to lack of space, we present only full replication (results are similar across all replication degrees).
% OLD: For every replication degree, the time grows nearly linearly with the number of points.
Figure~\ref{fig:eval:dssize_ingestion}(a) reports TREMOR's query answering time on 4 nodes with full replication as the dataset size (x-axis in log-scale) increases from 10\% to 100\% of the points of each dataset, i.e., from 4.2 to 42~GB for SeiFR (with $k=5$), and from 67 to 665~GB for Maule (with $t=0.9$).
%Since the cost of DMASS-V3 is $O(nm)$, i.e., linear in the length $n$ of the waveform for a fixed template length $m$, the figure also shows the time of DMASS-V3, scaled linearly from its time on the full datasets (Section~\ref{sec:eval:competitors}).
The time of TREMOR grows nearly linearly (slightly sub-linearly) with the dataset size, but much more slowly than DMASS-V3, with TREMOR's advantage growing with the dataset size.
On the full datasets, TREMOR takes 377s on Maule and 447~s on SeiFR, compared to 5,450s and 2,980s for DMASS-V3, i.e., it is 14.5$\times$ and 6.7$\times$ faster.
Due to lack of space, we present only full replication; the results are similar across all replication degrees. %, and for every replication degree, the time of TREMOR grows nearly linearly with the number of points.
%, since TREMOR answers every template with a scan or with FFTs, \tp{there is no index involved? 9a trela9w!} whose cost is proportional to the number of subsequences; replication only changes the slope. 
On Maule, the whole dataset (166.5 billion points) takes 633s without replication, 513s with Partial-2, and 377s with full replication (about 3.8, 3.1, and 2.3s per billion points); on SeiFR (10.5 billion points, with all 4,430 templates), it takes 483, 478, and 447s (about 46, 45, and 43s per billion points).
Overall, TREMOR is by far the only scalable solution for large seismic datasets.

\subsubsection{Index Ingestion Rate}
\label{sec:eval:ingestion}
\begin{figure}[t]
    \centering

    \begin{adjustbox}{max width=0.4\textwidth}
        \includegraphics{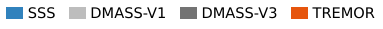}
    \end{adjustbox}

    % Row 1: varying threshold / k
    \begin{minipage}[t]{\columnwidth}
        \centering

        \begin{subfigure}[t]{0.49\textwidth}
            \centering
            \includegraphics[width=\textwidth]{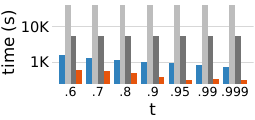}
            \vspace*{-0.6cm}
            \caption{Maule}
        \end{subfigure}
        \hfill
        \begin{subfigure}[t]{0.49\textwidth}
            \centering
            \includegraphics[width=\textwidth]{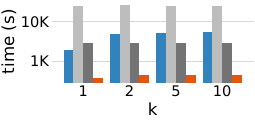}
            \vspace*{-0.6cm}
            \caption{SeiFR}
        \end{subfigure}

        \vspace{-0.3cm}
        \caption{Query answering time (log scale) for varying correlation threshold $t$ and $k$ (Full replication, 4 nodes).}
        \label{fig:eval:params}
    \end{minipage}

    \vspace{0.3cm}

    % Row 2: varying number of nodes
    \begin{minipage}[t]{\columnwidth}
        \centering

        \begin{subfigure}[t]{0.49\textwidth}
            \centering
            \includegraphics[width=\textwidth]{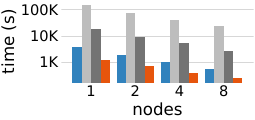}
            \vspace*{-0.6cm}
            \caption{Maule ($t=0.9$)}
        \end{subfigure}
        \hfill
        \begin{subfigure}[t]{0.49\textwidth}
            \centering
            \includegraphics[width=\textwidth]{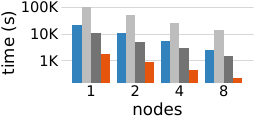}
            \vspace*{-0.6cm}
            \caption{SeiFR ($k=5$)}
        \end{subfigure}

        \vspace{-0.3cm}
        \caption{Query answering time (log scale) for varying number of nodes (Full replication).}
        \label{fig:eval:nodes}
    \end{minipage}

    \vspace{0.3cm}

    % Row 3: varying replication degree
    \begin{minipage}[t]{\columnwidth}
        \centering

        \begin{subfigure}[t]{0.49\textwidth}
            \centering
            \includegraphics[width=\textwidth]{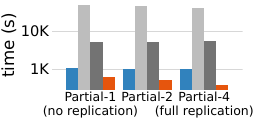}
            \vspace*{-0.5cm}
            \caption{Maule, $t=0.9$}
        \end{subfigure}
        \hfill
        \begin{subfigure}[t]{0.49\textwidth}
            \centering
            \includegraphics[width=\textwidth]{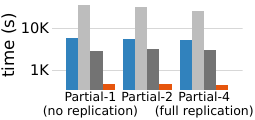}
            \vspace*{-0.5cm}
            \caption{SeiFR, $k=5$}
        \end{subfigure}

        \vspace{-0.3cm}
        \caption{Query answering time (log scale) for varying replication degree (4 nodes).}
        \label{fig:eval:repl}
    \end{minipage}

    \vspace{0.3cm}
    
    % Row 4
    \begin{minipage}[t]{\columnwidth}
        \centering
        \begin{adjustbox}{max width=0.75\textwidth}
            \includegraphics{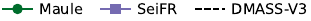}
        \end{adjustbox}
        
        \begin{subfigure}[t]{0.49\textwidth}
            \centering
            \includegraphics[width=\textwidth]{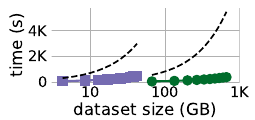}
            \vspace*{-0.4cm}
            \caption{Varying dataset size (log x-axis, 4 nodes, Full replication) 
           % \todo{x-axis: GBs} \tp{keep this graph around until we agree that the new one is better}
           }
        \end{subfigure}
        \hfill
        \begin{subfigure}[t]{0.49\textwidth}
            \centering
            \includegraphics[width=\textwidth]{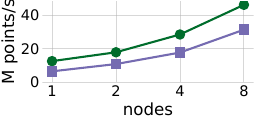}
            \vspace*{-0.4cm}
            \caption{Index Ingestion Rate (no replication)}
        \end{subfigure}

        \vspace{-0.2cm}
        \caption{Varying dataset size, and index ingestion rate. 
        \vspace{-0.4cm}
        }
        \label{fig:eval:dssize_ingestion}
    \end{minipage}
\end{figure}

Figure~\ref{fig:eval:dssize_ingestion}(b) shows the TREMOR ingestion rate (\# of waveform points indexed/second) on the largest channel of each dataset (1.82 and 1.58 billion points), without replication. 
From 1 to 8 nodes, the rate grows from 12.5 to 46.3 million points/second on Maule ($3.7\times$) and from 6.4 to 31.4 on SeiFR ($4.9\times$), since every node indexes only its part of the waveform.
SeiFR is ingested at one-half to two-thirds of the rate of Maule due to the larger subsequence sizes that SeiFR comes with (cf. Table~\ref{tab:eval:datasets}).
Without replication, indexing all channels takes 231 and 84 min on Maule on 1 and 8 nodes, and 27 and 5.9 min on SeiFR, respectively.
Replication lowers the ingestion rate, since every node indexes a larger part of the waveform.
With full replication, every node indexes the whole dataset, and the build takes longer than on a single node (290 vs. 231 min on Maule, 33 vs. 27 on SeiFR on 4 nodes).

\subsubsection{Ablation Study of TREMOR Parameters}
\label{sec:eval:parameters}
\label{sec:eval:ablation}
\label{sec:eval:sensitivity}
% OLD: TREMOR encapsulates a set of mechanisms to enable fast template matching (cf. Section~\ref{sec:tremor:adaptive}).
% OLD: We perform here an ablation study related to those mechanisms.
We study the mechanisms of TREMOR (Section~\ref{sec:tremor:adaptive}) through an ablation study.
In order to study the index related parameters, we use a highly selective configuration for Maule ($t=0.99$ and $t=0.999$) on three of its channels (150 templates), using 4 nodes, and full replication, without loss of generality.
We vary one parameter of TREMOR, while the others keep their fixed values (i.e., $f=1\%$, $b=5\%$, leaves of 2K subsequences, and 16 PAA segments).

\noindent \textbf{Scan fallback fraction $f$.}
TREMOR searches its index when at most a fraction $f$ of the sampled subsequences survive the lower bound, and scans or FFTs otherwise.
Never falling back to the scan (index only) is 9.3$\times$ slower at $t=0.99$, where the index cannot prune for most templates.
Among the values that we test (0, 0.1\%, 1\%, and 10\%), $f=1\%$ is the fastest at $t=0.99$ and 6\% slower than the fastest ($f=0.1\%$) at $t=0.999$.

\noindent \textbf{Lower-bound cut-off $b$.}
When TREMOR scans, it checks the lower bounds only if less than a fraction $b$ of the sampled subsequences pass them.
Checking the lower bounds pays off only when they prune almost every subsequence: at $t=0.99$, always checking them ($b=100\%$) is 1.6$\times$ slower than $b=5\%$ (6.7 vs.\ 4.1s).
Below 5\%, the cut-off matters less: $b=1\%$ takes up to 17\% less time.

\noindent \textbf{Leaf size.}
Leaves of 500 and 1K subsequences make the index construction 2.6$\times$ and 1.5$\times$ slower than leaves of 2K (427 and 240s vs.\ 162s), and the index 44\% and 25\% larger, without faster queries: the query time at $t=0.999$ is 2.0, 1.8, and 2.0s for leaves of 500, 1K, and 2K subsequences.
Leaves of 4K and 8K subsequences make the construction 1.5$\times$ faster (108 and 106s) for an index of the same size (within 3\%); the query time at $t=0.999$ is 1.9 and 2.0s.

\noindent \textbf{PAA segments.}
With 4 and 8 PAA segments instead of 16, the index is 55\% and 35\% smaller, but its lower bounds are looser: the query time at $t=0.999$ is 3.0 and 5.1s instead of 2.0s.

\vspace*{-0.2cm}
\subsection{Discussion}
Our deployments show that TREMOR changes the scale at which seismologists can perform template matching.
On 4 nodes, TREMOR is up to one order of magnitude faster than the current practice: % (DMASS-V3): 
it searches the 166.5 billion points of Maule for 11,922 templates in 5--10 minutes, and SeiFR for all 4,430 templates in 6--8 minutes, whereas DMASS-V3 requires about 1.5 hours and 50 minutes, and DMASS-V1 about 11 and 7 hours, respectively.
This is critical in practice, since seismologists repeat the search for different detection parameters, or when the template catalog is updated. 
For instance, the seven threshold searches over Maule take less than 50 minutes with TREMOR, compared to more than 10 hours with DMASS-V3 and more than 3 days with DMASS-V1.
Since TREMOR scales with the number of nodes, analyses that are currently restricted to small subsets of downsampled data, such as the SeiFR pipeline, can be performed over entire multi-year archives.
Moreover, the reference Maule analysis~\cite{flores2026fine} downsampled the waveforms to 25~Hz to keep the computation feasible. 
In contrast, TREMOR enables seismologists to analyze the data in their full detail, i.e., the original recordings at 50 and 100Hz, which translates to 4$\times$ more points. 

Pushing the envelope of what is computationally possible in practice is extremely important, especially for DAS, which records thousands of channels~\cite{zhan2020das} and relies on template matching to improve earthquake detection~\cite{li2018templatematching}.
We are now in the process of conducting more extensive seismological studies that produce new results, which we will report in a future article.

Overall, TREMOR allows seismologists to analyze entire multi-year archives and dense deployments at their original sampling rate, with large template catalogs and for multiple parameter settings, which was never before possible.

\section{Conclusions}
\label{sec:conclusion}
We presented TREMOR, a distributed framework for exact template matching on large seismic waveforms.
%TREMOR combines a distributed iSAX index with scan and FFT fallbacks, and chooses among them for every template.
% OLD: We deploy TREMOR in two real-world use cases in seismology, and through an extensive experimental evaluation we demonstrate its efficiency. 
% OLD: In particular, we show that it is up to 6.7$\times$ faster than the best competing method on each workload and up to 18$\times$ faster than \new{the fastest distributed version of} MASS, the FFT-based method popular in seismology, thus, enabling the analysts to execute their analysis pipelines \new{up to} one order of magnitude faster, as well as to analyze much larger, or much more detailed data, such as DAS recordings.
We deployed TREMOR on two real-world seismic use cases, where it is up to 6.7$\times$ faster than the best competing method on each workload, and up to 18$\times$ faster than the fastest distributed version of MASS, the FFT-based method popular in seismology. 
As a result, TREMOR enables seismologists to execute their analysis pipelines an order of magnitude faster, analyzing much larger, or much more detailed datasets. %, such as DAS recordings.

\begin{acks}
Work supported by EU Horizon project DataGEMS ($101188416$).
Manos Chatzakis is supported with a PhD  Scholarship from the Onassis Foundation.
Part of numerical computations were performed using the S-CAPAD/DANTE platform of IPGP.
We thank Azedine Mani, Head of the IT team (UFR Mathématiques et Informatique, Université Paris Cité), for his pivotal role in creating and making available computational resources essential to this work. 
\end{acks}

% \section*{Acknowledgments}
% Work supported by EU Horizon project DataGEMS (101188416).
% Manos Chatzakis is supported with a PhD scholarship from Onassis Foundation.

\bibliographystyle{ACM-Reference-Format}
\bibliography{references}

\end{document}